\documentclass[aip,reprint,jcp]{revtex4-1}
\usepackage[colorlinks=true,citecolor=blue]{hyperref}
\usepackage[pdftex]{graphicx}
\usepackage[english]{babel}
\usepackage{float}
\usepackage{amssymb}   
\usepackage{epsfig}

\begin{document}

\title{Accurate nonrelativistic ground-state energies of $3d$ transition metal atoms}

\author{A. Scemama, T. Applencourt, E. Giner, and M. Caffarel}
\affiliation{Lab. Chimie et Physique Quantiques, CNRS-Universit\'e de Toulouse, France.}

\begin{abstract}

We present accurate nonrelativistic ground-state energies of the transition metal atoms 
of the $3d$ series calculated with Fixed-Node Diffusion Monte Carlo (FN-DMC). 
Selected multi-determinantal expansions obtained with the CIPSI method (Configuration Interaction using a 
Perturbative Selection made Iteratively) and including the most prominent determinants of the full CI 
expansion are used as trial wavefunctions. Using a maximum of a few tens of thousands determinants, 
fixed-node errors on total DMC energies are found to be greatly reduced for some atoms with respect 
to those obtained with Hartree-Fock nodes. The FN-DMC/(CIPSI nodes) ground-state energies presented 
here are, to the best of our knowledge, the most accurate values reported so far. 
Thanks to the variational property of FN-DMC total energies, the results also provide lower bounds 
for the absolute value of all-electron correlation energies, $|E_c|$.
\end{abstract}

\maketitle

\section{Introduction}
\label{intro}
An accurate knowledge of nonrelativistic ground-state energies of atoms is 
known to be of great interest for computational chemistry. Atomic total energies are indeed routinely 
used to calibrate theoretical studies in electronic structure theory. For example, let us cite 
the search for more accurate exchange-correlation energy functionals in Density Functional Theory 
(DFT), the calibration of various approximations in wavefunction-based approaches 
(finite basis set effects, truncation at a given order in multi-particle excitations,
etc.), the study of the fixed-node approximation in quantum Monte Carlo 
(QMC), the definition of alternative/exotic electronic approaches, etc.
Furthermore, by combining experimental results and accurate nonrelativistic values, 
some valuable information about relativistic effects can also be obtained.

Here, accurate nonrelativistic
all-electron ground-state energies for the metal 
atoms of the $3d$ series (from Sc to Zn) are reported. Calculations are performed using the 
Fixed-Node Diffusion Monte Carlo (FN-DMC) approach, a quantum Monte Carlo (QMC) method known 
to be particularly powerful for computing ground-state energies.\cite{lestbook,rmp} 
An overwhelming number of works have been devoted to the calculation of 
accurate atomic ground-state energies using various highly-correlated approaches; 
here, we shall only restrict ourselves 
to briefly summarize the typical accuracies presently achievable. 
For small atoms (say, less than 10 electrons, that is, from H to Ne 
for neutral atoms) very accurate values with errors smaller than 10$^{-4}$--10$^{-5}$ a.u. 
(or much smaller for the ligthest atoms) can be obtained. For heavier atoms up to Ar (18 electrons), 
the accuracy reduces to the millihartree level ($\sim$ chemical
accuracy). For even bigger atoms (say, more than 20 electrons) to obtain a precision close 
the millihartree is problematic and only a small number of results have been published.
Regarding quantum Monte Carlo studies using FN-DMC or a closely related QMC variant,
most of the works have been concerned with atoms from Li to Ne; for the most recent ones, 
see {\it e.g.} [\onlinecite{brown1,buendia1,lopez,buendia2}]. For heavier atoms, 
most calculations have been performed using pseudo-potentials to remove core electrons 
(see {\it e.g.} [\onlinecite{mitas0}],[\onlinecite{wag1}], and [\onlinecite{luchow}]).
At the all-electron level, very little has been done. We can essentially 
cite the FN-DMC calculations by Ma {\it et al.}\cite{ma} for the Ar, Kr, and Xe atoms, 
calculations for the Cu atom and its cation,\cite{cu_caf,bouabca} and 
two studies by Buendia and collaborators for $3d$ transition metal atoms.\cite{buendia1,
buendia2}

It is fair to say that FN-DMC is presently the most accurate method 
for computing total ground state energies for large enough electronic systems. 
Potentially, diffusion Monte Carlo allows an exact stochastic solution of the Schr{\"o}dinger 
equation. Several sources of error make in practice FN-DMC simulations
non-exact. However, most of the errors are not of fundamental nature 
and can be easily kept under control (mainly, the statistical, finite time-step, 
and population control errors). In contrast, the fixed-node error resulting from the 
use of trial wavefunctions with 
approximate nodes is much more problematic since, up to now, 
no simple and systematic scheme to control this error has been devised. 
Note that the fixed-node approximation is variational, $E_{\rm FN} \ge E_0$, 
a convenient property to get upper and lower bounds for total energies
and absolute values of correlation energies, respectively [in contrast, {\it e.g.}, with the 
non-variational character of the commonly used CCSD(T) or M{\o}ller-Plesset approaches].

To decrease the fixed-node error, the common strategy is to use trial wavefunctions  
of the best possible quality and to resort to (large-scale) optimization 
techniques to get the best parameters entering the trial wavefunction (usually, 
via minimization of the total energy and/or its variance). A 
great variety of functional forms have been introduced for the wavefunction 
(see, {\it e.g.} [\onlinecite{mosko,flad,fili,braida,goddard,fili_vb,
bouabca,qmcgrid,mitas,sorella,rios}]),
and different optimization techniques designed to be efficient 
in a Monte Carlo context have been developed ({\it e.g.} [\onlinecite{cyrus}]). 
In this work, accurate nodes are built by employing a new class of trial wavefunctions 
very recently introduced in the context of QMC simulations.\cite{canadian} The wavefunction is expressed 
as a truncated Configuration Interaction (CI) expansion containing up to a few tens of 
thousands of determinants. The expansion is built thanks to the CIPSI method 
(Configuration Interaction using a Perturbative Selection made Iteratively).
The key point with CIPSI is the possibility of extracting the most prominent 
determinants of the FCI expansion. Very recent applications on several systems 
have shown that accurate nodes can be obtained.\cite{canadian,preprint} 
Furthermore, it has been observed that 
the quality of nodes appears to systematically improve when the number 
of determinants is increased. This property is remarkable since it allows 
a simple control of the fixed-node error.
Finally, an important practical property of CIPSI is that 
trial wavefunctions are generated in an automated way through the deterministic 
selection and diagonalization steps and the initial many-parameter 
stochastic optimization usually performed in QMC is avoided here.

The FN-DMC/(CIPSI nodes) total ground-state energies of metal atoms of the $3d$ series 
obtained here are compared to the very recent results of Buendia {\it et al.}\cite{buendia2}
An important and systematic improvement is obtained 
(lower total fixed-node energies). To the best of our knowledge, the data presented here are the best 
ones reported so far. Thanks to the variational property of FN-DMC total energies, 
the results also provide lower bounds for the absolute value of all-electron correlation energies, $|E_c|$.

\section{Methods and Computational details}
\label{methods}
\subsection{Configuration Interaction using a Perturbative Selection made Iteratively (CIPSI)}
\label{cipsi}
The CIPSI method, and similar approaches closely related, have been introduced 
and developed a long time ago by a number of authors (see, {\it e.g.}, 
[\onlinecite{bender,cipsi1,buenker1,buenker2,buenker3,bruna,buenker-book,cipsi2,harrison}]).
In a few words, the approach consists in building the multi-determinantal 
expansion {\it iteratively} by selecting at each step one determinant 
(or a group of determinants) according to a perturbative criterion. A 
determinant $D_i$ (or a group of determinants)
is added to the current wavefunction if its (their) energetic contribution(s) 
calculated by second-order perturbation theory is (are) sufficiently large.
In this way, the wavefunction is built hierarchically, 
the most important determinants of the FCI solution entering
first in the expansion. Such a construction must be contrasted with 
standard approaches (CIS, CISD, etc.) where the contributions at a given order 
are calculated by considering all possible particle-excitations 
with respect to some reference wavefunction (usually, the Hartree-Fock (HF) solution).
The CIPSI multi-determinantal expansion is thus much more compact than standard expansions,
an important practical point for FN-DMC where the trial wavefunction and 
its derivatives must be computed a very large number of times during the simulations.
Let us now briefly summarized the algorithm. More details can be found in Ref. [\onlinecite{canadian}] and 
in the original works cited above.

In multi-determinantal expansions the ground-state wavefunction $|\Psi_0\rangle$ 
is written as a linear combination of Slater determinants $\{|D_i\rangle\}$, 
each determinant corresponding to a given occupation by the 
$N_{\alpha}$ and $N_{\beta}$ electrons of $N=N_{\alpha}+N_{\beta}$ electrons
among a set of $M$ spin-orbitals $\{\phi_1,...,\phi_M\}$ (restricted case). 
The best representation of the exact wavefunction in the entire determinantal basis is the 
Full Configuration Interaction (FCI) wavefunction written as
\begin{equation}
|\Psi_0\rangle = \sum_i c_i |D_i\rangle
\end{equation}
where $c_i$ are the ground-state coefficients obtained by diagonalizing the Hamiltonian matrix, 
$H_{ij}=\langle D_i|H|D_j\rangle$, within the orthonormalized set, 
$\langle D_i|D_j\rangle= \delta_{ij}$, of determinants $|D_i\rangle$.\\

In its simplest form, the multi-determinant wavefunction is iteratively built as follows.
Let us call $|\Psi_0^{(n)}\rangle= \sum_{i \in S_n} c_i^{(n)} |D_i \rangle$ 
the current wavefunction at iteration $n$ where $S_n$ is the set of selected determinants at iteration $n$.
Typically, at the initial step $n=0$ a mono-determinantal HF-type or a short CAS-SCF-type wavefunction is used.
The first step consists in collecting all {\it different} determinants $|D_{i_c}\rangle$ 
connected by $H$ to $|\Psi_0^{(n)}\rangle$, that is $\langle \Psi_0^{(n)}|H|D_{i_c}\rangle \ne 0$.
Then, the second-order correction to the total energy resulting from each connected determinant is computed
\begin{equation}
\delta e(|D_{i_c}\rangle)=-\frac{{\langle \Psi_0^{(n)}|H|D_{i_c} \rangle}^2}{\langle D_{i_c}|H|D_{i_c}\rangle -E_0^{(n)}}
\label{e2pert}
\end{equation}
and the determinant (or group of determinants) $|D_{i_c^*}\rangle$ associated with the largest $|\delta e|$ 
(or greater than a given threshold) is (are) added to the reference subspace: 
$$S_{n} \rightarrow S_{n+1}= S_{n} \cup \{|D_{i_c^*} \rangle\}$$\\
Finally, the Hamiltonian matrix is then diagonalized within $S_{n+1}$ to obtain the new wavefunction 
at iteration $n+1$ and the process is iterated 
until a target size $N_{\rm dets}$ for the reference subspace is reached. The CIPSI wavefunction 
issued from this selection process is the trial wavefunction used here for FN-DMC. 

\subsection{Fixed-Node Diffusion Monte Carlo (FN-DMC)}
\label{qmc}
For a detailed presentation of the theoretical and practical aspects of FN-DMC, 
the reader is referred to the literature, {\it e.g} [\onlinecite{lester,towler,ency}]. 
Here, let us just 
emphasize that the central quantity of such approaches is the trial wavefunction $\Psi_T$ 
determining both the magnitude of the fixed-node error through its approximate 
nodes and the quality of the statistical convergence 
(good trial wavefunctions imply small statistical fluctuations).
The computational cost of FN-DMC is almost entirely determined by the evaluation 
at each Monte Carlo step of the value of $\Psi_T$ and its first (drift vector) and second 
derivatives (Laplacian needed for the local energy).
In view of the very large number of MC steps usually required 
(typically at least billions and often much more) 
it is essential
to be able of computing such quantities very rapidly.
In the present work, the typical size of the expansion considered 
is a few tens of thousands of determinants.
Some care is thus required when computing such expansions to keep the computational 
cost reasonable. The various aspects regarding this problem are 
presented in Ref. [\onlinecite{jcompchem}].

\subsection{Computational Details}
\label{details}

The atomic basis sets used for the calculations were the Slater-type 
orbitals of Bunge\cite{bunge1993roothaan} supplemented with four
additional $4f$
and three $5g$ functions (a total of 112 atomic basis functions).
All the CIPSI calculations were performed using Hartree-Fock molecular
orbitals using the code developed in our group (quantum package),
and all the FN-DMC calculations were performed using QMC=Chem.\cite{qmcchem}

For each atom, the CIPSI calculation was stopped when more than 
$10^6$ determinants were selected in the variational wave function.
This wave function was then truncated such that the least significant
determinants contributing to 0.5\% of the norm of the wave function
were discarded~: $10^4$--$5.10^4$ determinants were kept. This wave
function was used {\em without any modification} as the trial wave
function for the FN-DMC calculations (no Jastrow factor was used).

For the FN-DMC calculations, we have employed the algorithm described in
ref~[\onlinecite{PhysRevE.61.4566}] allowing us to use a small constant
number of walkers.
A block consisted in 30 walkers performing 5000 steps
with a time step of $10^{-5}$ a.u., a value chosen such that the
time-step error was smaller than the statistical error.
Long enough simulations have been performed 
to make the statistical error negligible with respect to the 
fixed-node one: depending on the atom, a number of blocks between
$7.10^4$ and $1.5\,10^5$ were calculated ($\sim 10^{10}$ MC steps).

\section{Results}
\label{results}
In table \ref{tab1} the variational energy, the number of determinants in the CIPSI expansion, 
and an estimate of the percentage of the total correlation energy (CE) recovered for each 
trial wavefunction $\Psi_T$ used in FN-DMC are given. The CE's reported 
are calculated from the recommended values given recently by McCarthy and Thakkar (denoted as McCT 
in what follows).\cite{thakkar} 
In sharp contrast with the present work, these values have not been computed directly 
from a unique (very) accurate energy calculation but have been obtained 
indirectly by combining M{\o}ller-Plesset (MP2) correlation energies  
extrapolated at the complete-basis-set (CBS) limit and CCSD(T) calculations
using Dunning's basis sets of various sizes. 
Note that the percentage of correlation energy 
already retrieved at the CIPSI variational level is around 60\%, a relatively important amount 
according to the standards of post-HF wavefunction theories for such systems.
In table \ref{tab2} the Fixed-Node DMC total energies obtained using standard Hartree-Fock nodes and
newly proposed CIPSI nodes are reported. For the sake of comparison, we also give the very recent 
results of Buendia {\it et al.}\cite{buendia2} that were up to now the lowest variational total energies 
reported for these atoms. In their study the trial wavefunctions employed 
are written as the product of a nodeless correlation factor and a so-called model function obtained 
within the parametrized Optimized Effective Potential (OEP) approximation. The model function determining 
the nodal structure is written as a linear combination of a few Configuration State Functions (CSFs), 
mainly to take into account $4s-4p$ near-degeneracy effects. For the Cr and Cu atoms with
a singly occupied $4s$ shell the model function is represented by a single CSF, while for the other 
atoms $4s^2$$3d^n$ and $4p^2$$3d^n$ configurations are mixed. For each type of nodes used, 
an estimate of the percentage of the correlation energy is also reported. The percentages retrieved 
by all FN-DMC calculations presented are important and range between 89 and 94\%. 
A first observation is that energies resulting from HF and OEP nodes are of comparable quality, 
while CIPSI nodes may lead to significantly lower fixed-node energies. The gain in energy with the 
new nodes is found to decrease almost uniformly with $Z$. For the lightest elements (Sc, V and Ti)
a maximum gain of about 0.04 a.u is achieved; for the intermediate atoms (Cr to Ni) 
about 0.02-0.03 a.u. is obtained, while for the two heaviest elements (Cu and Zn) no gain is 
observed within statistical fluctuations. The fact that CIPSI performs better 
for lighest elements is not surprising since Hartree-Fock nodes are known to be
well-adapted to atoms with spherical symmetry. In the extreme case of the Cu and Zn atoms 
having a totally filled and spherically symmetric $3d$ shell, HF and CIPSI nodes give similar results. 
In the opposite case of light atoms, the CIPSI wavefunctions, that have many more degrees 
of freedom than the single-configuration HF solution to describe 
non-symmetrical electronic configurations, lead to much improved results.
In table \ref{tab3} the correlation energies resulting from our FN-DMC simulations are reported and 
compared to the recommended values of McCT.
As already noted, these latter results have been obtained with a mixed approach including 
MP2-CBS and CCSD(T) calculations. According to the authors, the errors in these values are estimated 
to be $\pm 3\%$. The relative differences between FN-DMC/[HF nodes] or FN-DMC/[OEP nodes] and 
the McCT values go from 8 to 11\%. Using CIPSI nodes the differences are reduced 
and range between 6 and 8\%. Note that the typical statistical error on these percentages 
is small and about 0.2\%.
Although our final values for correlation energies are slightly less accurate than the estimations 
made by McCT, we would like emphasize and conclude on three important points:
i.) In contrast with what has been done by McCT, our correlation energies have been 
directly computed with a unique highly-correlated electronic structure method. No hybrid scheme mixing 
results of two different approaches has been employed. To the best of our knowledge, the FN-DMC 
values presented here are the most accurate (lowest) nonrelativistic total energies 
ever reported for the $3d$ transition metal atoms.
ii.) As a consequence of the variational property of FN-DMC total energies and, also in contrast 
with McCT's results, the absolute values of our correlation energies are {\it exact lower bounds} 
of the unknown CE's.
iii) Finally, in view of the great versatility of FN-DMC/CIPSI, there is no reason why 
improved lower bounds would not be achieved in the near future, 
thus leading to benchmark-type results for such atoms.
\begin{table}[t]
\begin{center}
\begin{tabular}{|l|c|c|c|c|c|c|}
\hline
Atom &  E$_{\rm var}$(CIPSI)     & [CE in \%] & N$_{\rm dets}$\\ 
\hline               
Sc  &  -760.32556 & [66.5\%]  &     11 389        \\ 
Ti  &  -849.02624 & [66.9\%]  &     14 054        \\ 
V   &  -943.53667 & [64.9\%]  &     12 441        \\ 
Cr  & -1044.03692 & [63.6\%]  &     10 630        \\ 
Mn  & -1150.57902 & [63.0\%]  &     11 688        \\ 
Fe  & -1263.21805 & [62.5\%]  &     13 171       \\ 
Co  & -1382.24964 & [62.8\%]  &     15 949        \\ 
Ni  & -1507.74694 & [62.3\%]  &     15 710        \\ 
Cu  & -1639.96605 & [63.3\%]  &     48 347        \\ 
Zn  & -1778.82784 & [60.5\%]  &     44 206       \\ 
\hline
\end{tabular}
\end{center}
\caption{Variational energy, E$_{\rm var}$(CIPSI), of the CIPSI trial wavefunctions $\Psi_T$ used in FN-DMC, 
estimated percentages of the correlation energy (CE) recovered, and
number of determinants, N$_{\rm dets}$, in the expansions.
Energy in hartree.}
\label{tab1}
\end{table}

\begin{table*}[t]
\begin{center}
\begin{tabular}{|l|rl|rl|rl|c|}
\hline
Atom$^a$& HF nodes & [CE in \%]  & OEP nodes$^b$ & [CE in \%]   & 
CIPSI nodes & [CE in \%]  & FN energy gain with CIPSI nodes$^c$\\ 
\hline                                                                                         
Sc $[s^2 d^1]$  &  -760.5265(13)  & [89.2(2)\%]   &             -760.5288(6)            &  [89.50(7)\%]  &   -760.5718(16)    & [94.4(2)\%]  &    -0.0453(21)   \\ 
Ti $[s^2 d^2]$ &  -849.2405(14)  & [89.6(2)\%]   &             -849.2492(7)            &  [90.55(7)\%]  &   -849.2841(19)    & [94.2(2)\%]  &    -0.0436(24)   \\ 
V $[s^2 d^3]$  &  -943.7843(13)  & [89.6(1)\%]   &             -943.7882(6)            &  [89.95(6)\%]  &   -943.8234(16)    & [93.4(2)\%]  &    -0.0391(21)   \\ 
Cr $[s^1 d^5]$ &  -1044.3292(16) & [91.0(2)\%]  &            -1044.3289(6)            &   [90.93(6)\%]  &   -1044.3603(17)   & [93.9(2)\%] &    -0.0311(23)   \\ 
Mn $[s^2 d^5]$  &  -1150.8880(17) & [90.4(2)\%]   &            -1150.8897(7)            &  [90.54(6)\%]  &   -1150.9158(20)   & [92.9(2)\%]  &    -0.0278(26)   \\ 
Fe $[s^2 d^6]$ &  -1263.5589(19) & [90.1(2)\%]   &            -1263.5607(6)            &  [90.26(5)\%]  &   -1263.5868(21)   & [92.4(2)\%]  &    -0.0279(28)   \\ 
Co $[s^2 d^7]$  &  -1382.6177(21) & [90.5(2)\%]   &            -1382.6216(8)            &  [90.85(6)\%]  &   -1382.6377(24)   & [92.1(2)\%]  &    -0.0200(32)   \\ 
Ni $[s^2 d^8]$  &  -1508.1645(23) & [91.6(2)\%]   &            -1508.1743(7)            &  [92.27(5)\%]  &   -1508.1901(25)   & [93.4(2)\%]  &    -0.0256(34)   \\ 
Cu $[s^1 d^{10}]$ &  -1640.4271(26) & [92.4(2)\%]   &            -1640.4266(7)            &  [92.34(4)\%]  &   -1640.4328(29)   & [92.7(2)\%]  &    -0.0057(39)   \\ 
Zn $[s^2 d^{10}]$ &  -1779.3371(26) & [91.9(2)\%]   &            -1779.3425(8)            &  [92.24(5)\%]  &   -1779.3386(31)   & [92.0(2)\%]  &    -0.0015(40)   \\ 
\hline
\end{tabular}
\end{center}
\caption{FN-DMC total energies for the $3d$ series of transition metal atoms together with the percentage 
of correlation energy recovered for different nodal structures. Energy in hartree.\\
$^a$ Atom given with its electronic configuration, the common argon core 
[Ar]=(1s$^2$2s$^2$2p$^6$3s$^2$3p$^6$) being not shown.\\
$^b$ Ref. [\onlinecite{buendia2}].\\
$^c$ Difference between FN-DMC energies obtained with HF nodes (column 1) and newly proposed CIPSI nodes (column 3).}
\label{tab2}
\noindent
\end{table*}

\begin{table}[t]
\begin{center}
\begin{tabular}{|l|c|c|c|c|}
\hline
Atom&HF nodes&OEP nodes$^a$&CIPSI nodes& McCT$^b$\\
\hline
Sc  & 0.7900(13) & 0.7923(6)&0.8353(16) &0.8853\\
Ti  & 0.8454(14) & 0.8541(7)&0.8890(19) &0.9433\\
V   & 0.9000(13) & 0.9039(6)&0.9390(16) &1.0049\\
Cr  & 0.9728(16) & 0.9725(6)&1.0039(17) &1.0695\\
Mn  & 1.0218(17) & 1.0235(7)&1.0495(20) &1.1304\\
Fe  & 1.1122(19) & 1.1140(6)&1.1401(21) &1.2343\\
Co  & 1.2016(21) & 1.2055(8)&1.2216(24) &1.3270\\
Ni  & 1.3043(23) & 1.3141(7)&1.3299(25) &1.4242\\
Cu  & 1.4634(26) & 1.4629(7)&1.4691(29) &1.5842\\
Zn  & 1.4890(26) & 1.4944(8)&1.4905(31) &1.6202\\
\hline
\end{tabular}
\end{center}
\caption{Fixed-Node DMC correlation energies,$-E_c$, in hartree using HF and 
CIPSI nodes. Comparison with the recommended values of McCarthy and Thakkar (McCT).\cite{thakkar} \\
$^a$ Ref. [\onlinecite{buendia2}].\\
$^b$ Ref. [\onlinecite{thakkar}]}
\label{tab3}
\end{table}

{\it Acknowledgments.}
AS and MC thank the Agence Nationale pour la Recherche (ANR) for support 
through Grant No ANR 2011 BS08 004 01.
This work has been possible thanks to the computational support of CALMIP (Toulouse) through a Meso-Challenge on their new Eos supercomputer
(\url{http://www.calmip.univ-toulouse.fr}).

\bibliography{atoms}{}

\begin{thebibliography}{10}

\bibitem{lestbook}
B.L. Hammond, W.A.~Lester Jr., and P.J. Reynolds.
\newblock {\em Monte Carlo Methods in Ab Initio Quantum Chemistry}, volume~1 of
  {\em World Scientific Lecture and Course Notes in Chemistry}.
\newblock 1994.

\bibitem{rmp}
W.M.C. Foulkes, L.~Mitas, R.G. Needs, and G.~Rajagopal.
\newblock {\em Rev. Mod. Phys.}, 73:33, 2001.

\bibitem{brown1}
M.D. Brown, J.R. Trail, P.~Lopez Rios, and R.J. Needs.
\newblock {\em J. Chem. Phys.}, 126:224110, 2007.

\bibitem{buendia1}
A.~Sarsa, E.Buendia, F.J. Galvez, and P.~Maldonado.
\newblock {\em J. Phys. Chem. A}, 112:2074, 2008.

\bibitem{lopez}
P.~Lopez Rios, P.~Seth, N.D. Drummond, and R.J. Needs.
\newblock {\em Phys. Rev. E}, 86:036703, 2012.

\bibitem{buendia2}
E.Buendia, F.J. Galvez, P.~Maldonado, and A.~Sarsa.
\newblock {\em Chem. Phys. Lett.}, 559:12, 2012.

\bibitem{mitas0}
L.~Mitas.
\newblock {\em Phys. Rev. A}, 49:4411, 1994.

\bibitem{wag1}
L.~Wagner and M.~Mitas.
\newblock {\em Chem. Phys. Lett.}, 370:412, 2003.

\bibitem{luchow}
S.~Sokolova and A.~Luchow.
\newblock {\em Chem. Phys. Lett.}, 320:421, 2000.

\bibitem{ma}
A.~Ma, N.D. Drummond, M.D. Towler, and R.J. Needs.
\newblock {\em Phys. Rev. E}, 71:066704, 2005.

\bibitem{cu_caf}
M.~Caffarel, J.P. Daudey, J.L Heully, and A.~Ramirez-Solis.
\newblock {\em J. Chem. Phys.}, 123:094102, 2005.

\bibitem{bouabca}
T.~Bouab\c{c}a, B.~Bra\^{\i}da, and M.~Caffarel.
\newblock {\em J. Chem. Phys.}, 133:044111, 2010.

\bibitem{mosko}
K.E. Schmidt and J.W. Moskowitz.
\newblock {\em J. Chem. Phys.}, 93:4172, 1990.

\bibitem{flad}
H.J. Flad, M.~Caffarel, and A.~Savin.
\newblock {\em Quantum Monte Carlo calculations with multi-reference trial wave
  functions}.
\newblock Recent Advances in Quantum Monte Carlo Methods. World Scientific
  Publishing, 1997.

\bibitem{fili}
C.~Filippi and C.J. Umrigar.
\newblock {\em J. Chem. Phys.}, 105:213, 1996.

\bibitem{braida}
B.~Bra\i{i}da, J.~Toulouse, M.~Caffarel, and C.~J. Umrigar.
\newblock {\em J. Chem. Phys.}, 134:0184108, 2011.

\bibitem{goddard}
A.~G. Anderson and W.A.~Goddard III.
\newblock {\em J. Chem. Phys.}, 132:164110, 2010.

\bibitem{fili_vb}
F.~Fracchia, C.~Filippi, and C.~Amovilli.
\newblock {\em J. Chem. Theory Comput.}, 8:1943, 2012.

\bibitem{qmcgrid}
A.~Monari, A.~Scemama, and M.~Caffarel.
\newblock Large-scale quantum monte carlo electronic structure calculations on
  the egee grid.
\newblock In Franco Davoli, Marcin Lawenda, Norbert Meyer, Roberto Pugliese,
  Jan Węglarz, and Sandro Zappatore, editors, {\em Remote Instrumentation for
  eScience and Related Aspects}, pages 195--207. Springer New York, 2012.
\newblock http://dx.doi.org/10.1007/978-1-4614-0508-5\_13.

\bibitem{mitas}
M.~Bajdich, L.~Mit\'a\v{s}, G.~Drobn\`y, and L.K. Wagner.
\newblock {\em Phys. Rev. Lett.}, 96:240402, 2006.

\bibitem{sorella}
M.~Casula, C.~Attaccalite, and S.~Sorella.
\newblock {\em J. Chem. Phys.}, 121:7110, 2004.

\bibitem{rios}
P.~Lopez Rios, A.~Ma, N.D. Drummond, M.D. Towler, and R.J. Needs.
\newblock {\em Phys. Rev. E}, 74:066701, 2006.

\bibitem{cyrus}
C.J. Umrigar, J.~Toulouse, C.~Filippi, S.~Sorella, and R.G. Hennig.
\newblock {\em Phys. Rev. Lett.}, 98:110201, 2007.

\bibitem{canadian}
E.~Giner, A.~Scemama, and M.~Caffarel.
\newblock {\em Can. J. Chem.}, 91:879, 2013.

\bibitem{preprint}
E.~Giner, A.~Scemama, and M.~Caffarel.
\newblock 2014.

\bibitem{bender}
C.~F. Bender and E.~R. Davidson.
\newblock {\em Phys. Rev.}, 183:23, 1969.

\bibitem{cipsi1}
B.~Huron, P.~Rancurel, and J.P. Malrieu.
\newblock {\em J. Chem. Phys.}, 58:5745, 1973.

\bibitem{buenker1}
R.~J. Buenker and S.~D. Peyerimholf.
\newblock {\em Theor. Chim. Acta}, 35:33, 1974.

\bibitem{buenker2}
R.~J. Buenker and S.~D. Peyerimholf.
\newblock {\em Theor. Chim. Acta}, 39:217, 1975.

\bibitem{buenker3}
R.~J. Buenker, S.~D. Peyerimholf, and W.~Butscher.
\newblock {\em Mol. Phys.}, 35:771, 1978.

\bibitem{bruna}
P.~J. Bruna, D.~S. Peyerimholf, and R.~J. Buenker.
\newblock {\em Chem. Phys. Lett.}, 72:278, 1980.

\bibitem{buenker-book}
R.~J. Buenker, S.~D. Peyerimholf, and P.~J. Bruna.
\newblock {\em Computational Theoretical Organic Chemistry}.
\newblock Reidel, Dordrecht, 1981.

\bibitem{cipsi2}
S.~Evangelisti, J.P. Daudey, and J.P. Malrieu.
\newblock {\em Chem. Phys.}, 75:91, 1983.

\bibitem{harrison}
R.~J. Harrison.
\newblock {\em J. Chem. Phys.}, 94:5021, 1991.

\bibitem{lester}
B.~L. Hammond, W.A.~Lester Jr., and P.J. Reynolds.
\newblock {\em Monte Carlo Methods in Ab Initio Quantum Chemistry}, volume~1 of
  {\em Lecture and Course Notes in Chemistry}.
\newblock World Scientific, Singapore, 1994.
\newblock World Scientific Lecture and Course Notes in Chemistry Vol.1.

\bibitem{towler}
M.D. Towler.
\newblock {\em Quantum Monte Carlo, or, how to solve the many-particle
  Schrödinger equation accurately whilst retaining favourable scaling with
  system size}.
\newblock Wiley, 2011.

\bibitem{ency}
M.~Caffarel.
\newblock {\em Quantum Monte Carlo Methods in Chemistry}.
\newblock Encyclopedia of Applied and Computational Mathematics. Springer,
  2012.

\bibitem{jcompchem}
E.Buendia, F.J. Galvez, P.~Maldonado, and A.~Sarsa.
\newblock {\em J. Comp. Chem.}, 34:938, 2013.

\bibitem{bunge1993roothaan}
C.~F. Bunge, J.A. Barrientos, and A.~Bunge-Vivier.
\newblock Roothaan-hartree-fock ground-state atomic wave functions: Slater-type
  orbital expansions and expectation values for z= 2-54.
\newblock {\em Atomic data and nuclear data tables}, 53(1):113--162, 1993.

\bibitem{qmcchem}
See web site: "Quantum Monte Carlo for Chemistry@Toulouse",
  http://qmcchem.ups-tlse.fr.

\bibitem{PhysRevE.61.4566}
Roland Assaraf, Michel Caffarel, and Anatole Khelif.
\newblock Diffusion monte carlo methods with a fixed number of walkers.
\newblock {\em Phys. Rev. E}, 61(4):4566--4575, Apr 2000.

\bibitem{thakkar}
S.P. McCarthy and Thakkar A.J.
\newblock {\em J. Chem. Phys.}, 136:054107, 2012.

\end{thebibliography}
\bibliographystyle{unsrt}
\end{document}